\documentclass[prd,onecolumn,floats,floatfix,showpacs,nofootinbib, 12pt]{revtex4}
\usepackage{graphicx}
\usepackage{dcolumn}
\usepackage{bm}
\usepackage{graphics}
\usepackage{slashed}
\usepackage{amssymb}
\usepackage{natbib}
\usepackage{amsmath}
\usepackage{url}
\usepackage{color}
\newcommand{\bea}{\begin{eqnarray}}
\newcommand{\ena}{\end{eqnarray}}
\newcommand{\bean}{\begin{eqnarray*}}
\newcommand{\enan}{\end{eqnarray*}}

\newtheorem{theorem}{Theorem}

\begin{document}

\title{Hidden geometries in nonlinear theories: \\a novel aspect of analogue gravity}

\author{E.Goulart\footnote{egoulart@cbpf.br}, 
M. Novello\footnote{M. Novello is Cesare Lattes ICRANet Professor: novello@cbpf.br}, F.T.Falciano\footnote{ftovar@cbpf.br}, J.D. Toniato\footnote{toniato@cbpf.br}}

\affiliation{Instituto de Cosmologia Relatividade Astrofisica ICRA -
CBPF\\ Rua Dr. Xavier Sigaud, 150, CEP 22290-180, Rio de Janeiro,
Brazil}

\date{\today}

\begin{abstract}
We show that non-linear dynamics of a scalar field $ \varphi$ may be described as a modification of the spacetime geometry. Thus, the self-interaction is interpreted as a coupling of the scalar field with an effective gravitational metric that is constructed with $\varphi$ itself.
We prove that this process is universal, that is, it is valid for arbitrary Lagrangian. Our results are compared to usual analogue models of gravitation, where the emergence of a metric appears as a consequence of linear perturbation.
\end{abstract}

\pacs{02.40.Ky, 04.20.Cv, 04.20.Gz}
 \maketitle

\section{Introduction}

Analogue gravity has become an active arena of relativistic physics in recent years.  This terminology involves the description of distinct physical processes in terms of an effective modification of the metrical structure of a background space-time. The basic idea is to investigate aspects of general relativity using systems that may be reproduced in the laboratory or admit a simple geometrical interpretation of their physical features. The analogies may encompass classical and quantum aspects of  fields in curved spacetimes and has been concentrated in studying artificial black holes, emergent spacetimes, effective signature transitions, breakdown of Lorentz invariance and quantum gravity phenomenology ( a complete list of references can be found in \cite{viss} and \cite{abh}).

Until now, the analogies have focused only into perturbative aspects of the system. Hence, it is restricted mainly to the propagation of excitations (photons or quasi-particles) through a given background configuration \cite{Un}-\cite{E}. In this way, the relevant equations describe an approximative solution that consider linearized fluctuations over a given background configuration. The evolution of these perturbations is always governed by an effective metric which could be associated to a specific gravitational configurations. By ``specific" it is understood that only aspects of \textit{test} fields propagating in a \textit{given}  gravitational background are considered i.e all effects due to gravitational back-reaction should be negligible (see \cite{Sil} for a pedagogical exposition). We note that in this scheme only perturbations ``perceive'' the effective metric. 

However, the effective metric is not a typical and exclusive component of perturbative phenomena, intimately associated to linearization on top of a background. Indeed, as we shall prove in this paper, we claim that it is possible to describe the dynamics of scalar field $\varphi$ in terms of an emergent geometrical configuration. Then we may interpret the equation of motion as if $\varphi$ is embedded in an effective curved structure generated by itself. This accomplish a new  geometrization scheme for the dynamics of $\varphi$.

In a recent communication \cite{novello}, two of us have shown that it is possible to go beyond some of the linearized approximations in the case of a scalar field. This means that there exist a special situation such that the nonlinear equation of motion of $\varphi$ can be described equivalently as a field propagating in a curved spacetime. Our previous result was restricted only to a unique lagrangian, which was given in terms of an infinite series of powers of $w$, where $w=\gamma^{\mu\nu}\partial_{\mu}\varphi\partial_{\nu}\varphi$. 

In the present paper we give a step ahead and show that our previous result is much broader and constitutes a general property of any self-interacting relativistic scalar field. Our fundamental claim may be summarized as: 
\begin{itemize}
\item{The dynamics of a relativistic scalar field endowed with a lagrangian $L(w,\varphi)$ can be described as if $\varphi$ interacts minimally  with an emergent metric constructed solely in terms of $\varphi$ and its derivatives.}
\end{itemize}
It is important to emphasize that our result is completely independent of any process of linearization and does not rely on any kind of approximation. In the particular case of excitations on top of a given background solution, our scheme furnishes a recipe to re-obtain all the usual results typical to analogue models after a straightforward linearization procedure. In addition, we discuss how the back-reaction issue is connected to our results.

In the next section, we summarize the common knowledge in analogue models stressing some of the points that might help to distinguish our result. In section \ref{geomfull} we develop the geometrization of the dynamics of any scalar non-linear field. Section \ref{abri} is devoted to analyze the back-reaction issue, while in section \ref{ghf} we connect our result with the description of hydrodynamical fluids.

\section{Effective geometries: a brief review}

In this section we briefly review some well known results and approximations concerning the effective metric technique.  For the sake of simplicity, let us consider a relativistic real scalar field $\varphi$ propagating in a flat Minkowski spacetime with a nonlinear dynamics provided by the action \cite{bar1}
\[
S=\int L(\varphi, w) \sqrt{-\gamma}\  d^{4}x
\]
where $w \equiv \gamma^{\mu\nu} \partial_{\mu} \varphi \, \partial_{\nu} \varphi$ is the canonical kinetic term and $\gamma=det(\gamma_{\mu\nu})$ is the determinant of the metric in arbitrary curvilinear coordinates. The equation of motion immediately reads
\begin{equation}\label{EqMotion}
\Bigl(\ L_{w} \, \partial_\nu \varphi \, \gamma^{\mu\nu} \Bigr)_{;\mu} = \frac{1}{2}L_{\varphi}
\end{equation}
where $L_X$ denotes the first derivative of $L$ with respect to the variable $X$ and ``;'' means covariant derivative with respect to $\gamma_{\mu\nu}$. This is a quasi-linear second order partial differential equation for $\varphi$ of the form \cite{cou}
\begin{equation}\label{mot}
\hat{g}^{\alpha\beta}\left(x,\varphi,\partial\varphi\right)\partial_{\alpha}\partial_{\beta}\varphi+ F\left(x,\varphi,\partial\varphi\right)=0,
\end{equation}
where $F\left(x,\varphi,\partial\varphi\right)$ stands for terms depending only on the curvilinear coordinates $x$, the field $\varphi$ and its first derivatives $\partial\varphi$. A straightforward calculation shows that the object $\hat{g}^{\alpha \beta}$ can be expressed as 
\begin{equation}\label{met}
\hat{g}^{\mu\nu}\equiv L_{w}\gamma^{\mu\nu}+2L_{ww}\partial^{\mu}\varphi\partial^{\nu}\varphi
\end{equation}
and determines the principal part of the equation of motion i.e. the part that involves the higher order derivative terms. In what follows we discuss how the $\hat{g}^{\alpha\beta}$ may be associated to the contravariant components of a riemannian effective spacetime. Let us note that there exists two complementary aspects of this association. In a typical condensed matter system, these aspects may be identified respectively with the geometrical and physical regimes of acoustics.

\subsection{Ray propagation}\label{rayprop}

The first aspect that may be rephrased in terms of an effective spacetime occurs in the realm of geometrical optics approximation. Although the main part of this discussion may be elegantly stated in terms of Hadamard's formalism of discontinuities \cite{Had,Had1}, we develop our arguments in the context of the eikonal approximation. The aim of the approximation is to evaluate the characteristic surfaces of the nonlinear equation (\ref{EqMotion}). The basic idea is to consider a continuous solution $\varphi_{0}$ of (\ref{EqMotion}) and a family of approximated wavelike solutions of the form \cite{Volker}
\begin{equation}
 \varphi(x)=\varphi_{0}(x)+\alpha f(x)exp(iS(x)/\alpha),
\end{equation}
where $\alpha$ is a real parameter and both the amplitude $f(x)$ and the phase $S(x)$ are continuous functions. As long as, by assumption, both $\varphi(x)$ and $\varphi_0(x)$ satisfy (\ref{EqMotion}), in the limit of a rapidly varying phase, which is equivalent to take $\alpha \rightarrow 0$, we find the dispersion relation
\begin{equation}\label{1}
\hat{g}^{\alpha\beta}k_{\alpha}k_{\beta}=0
\end{equation}
with $k_{\mu}\equiv S_{,\mu}$ and $\hat{g}^{\alpha\beta}$ being evaluated at the solution $\varphi_{0}$. This is the eikonal equation, it constitutes a first order nonlinear PDE for $S(x)$ and determines the causal structure of the theory \cite{tan}. Now, suppose that the matrix $\hat{g}^{\alpha\beta}$ is invertible, i.e. there exist $\hat{g}_{\mu \nu}$ such that $\hat{g}^{\mu \alpha}\hat{g}_{\alpha \nu}=\delta^\mu_{ \;\nu}$. Defining its covariant derivative such that $\hat{g}_{\alpha\beta ||\nu}$=0, we obtain that, once the
vector $ k_{\mu}$ is a gradient, the following equation holds  
\begin{equation}\label{2}
\hat{g}^{\mu\nu}k_{\alpha ||\mu}k_{\nu}=0\quad .
\end{equation}
The above result allows one to interpret the rays describing the perturbations of the scalar field as if they were propagating as null geodesics in the effective metric $\hat{g}_{\mu\nu}$ i.e. the effective metric determines the causal structure of field excitations. Thus, there exist two distinct metrics in this framework: the background Minkowskian $\gamma^{\mu\nu} $ that enters in the dynamics of the field $\varphi$  and the effective metric $ \hat{g}^{\mu\nu}$ that controls the propagation of rays in the geometrical optics limit.  

Given that both equations (\ref{1}) and (\ref{2}) are invariant under conformal transformations, we obtain that $k_{\alpha}$ is a null geodesic with respect to any metric proportional to the effective metric as it was defined in (\ref{met}). Thus, there is a degeneracy of metrics in the sense that any metric conformally related to $\hat{g}_{\mu \nu}$ equally describes the evolution of rays. One may conveniently choose one of the above metrics  to investigate ray propagation in curved spacetimes. This conformal freedom works only for the perturbations in the geometrical optics limit. As we will see below, it cannot be implemented in the next aspect of analogue gravity, namely, when we consider wave-like propagation.

\subsection{Wave propagation}\label{waveprop}

The second type of analogy with gravitational physics comes from a relaxation of the geometrical optics limit. In other words, we shall look for the dynamics of an arbitrary first order field perturbations $\delta\varphi$ and show that they satisfy a wave-like equation in an effective curved manifold. To do so, we again consider a continuous solution $\varphi_{0}$ of the equation (\ref{EqMotion}) and seek for the equation that governs the evolution of the perturbations around this background solution, i.e
\begin{equation}\label{pertu} 
\varphi=\varphi_{0}+\delta\varphi \qquad \mbox{with}\quad \delta \varphi^{2} \ll \delta\varphi \qquad.
\end{equation}

As has been shown sometimes in the literature (see, for instance, \cite{waveprop}, \cite{meta}), a straightforward calculation yields a Klein-Gordon-like equation in an effective spacetime whose metric $\hat{f}^{\mu\nu}$ is determined by the background configuration $\varphi_{0}$. Inserting the above ansatz in (\ref{EqMotion}) and keeping only terms up to first order in $\delta \varphi$, we can recast the equation of motion for the perturbation as
\begin{equation}\label{wavv}
\square_{\hat{f}}\ \delta\varphi+m_{eff}^{2}\delta\varphi=0
\end{equation}
where we defined the effective metric $\hat{f}^{\mu \nu}$ and the effective mass term $m_{eff}$ as
\begin{eqnarray}
\hat{f}^{\mu\nu}&=&L_{w}^{-2}(1+\beta w)^{-1/2}\hat{g}^{\mu\nu}\quad , \qquad \mbox{with} \quad \qquad \beta\equiv 2L_{ww}/L_{w}\\
&&\nonumber\\
m^{2}_{eff}&\equiv& L_{w}^{-2}(1+\beta w)^{-1/2}\left[L_{\varphi\varphi w}w-\frac{1}{2}L_{\varphi\varphi}+\frac{\partial\hat{g}^{\alpha\beta}}{\partial\varphi}\varphi_{,\alpha ;\beta}\right]\quad .
\end{eqnarray}
Inasmuch we are dealing with more than one metric, we though it would be conveniently to introduce a notation to specify with which metric tensor the d'Alembertian is constructed, i.e. we define the notation
\begin{equation}\label{dal}
\square_{\hat{f}}\ \delta\varphi\equiv\frac{1}{\sqrt{-\hat{f}}}(\sqrt{-\hat{f}}\hat{f}^{\mu\nu}\delta\varphi_{,\mu})_{,\nu}\quad .
\end{equation}

Note that both the effective metric and the effective mass should be evaluated with the background solution $\varphi_{0}(x)$, leading to a linear equation for the perturbation $\delta\varphi$. Thus, the perturbation propagates as a massive scalar field in an effective emergent spacetime that makes no reference to the perturbation itself.  We stress that this result  is valid for all sufficiently small excitation and has nothing to do with the frequency of the wave. Both sections \ref{rayprop} and \ref{waveprop} deal with approximations. Nevertheless they are complementary in the sense that in the optical regime one considers only waves with small amplitudes and very short wave-length (very high frequency) while in the other the amplitude is made very small letting the frequency be completely arbitrarily.

\section{Geometrization of Field Dynamics}\label{geomfull}

So far we have treated only perturbative aspects of propagation. From now on we are going to investigate the relation between the full equation of motion (\ref{EqMotion}) and the effective metric seen by its excitations. Thus, we want to address the question of whether it is possible that both the perturbation and the field itself propagate in a similar emergent scenario. Note that this is exactly the case for linear theories, i.e. for linear scalar field theory both the field and its excitations propagate in the same background metric, namely the Minkowskian spacetime. 

A non-trivial situation appears when one considers non-linear theories. In \cite{novello} it was shown that for a non-linear theory given by a specific lagrangian the dynamics of the scalar field and its perturbations can be described as if they both were immersed in an effective curved spacetime. Thus, as it happens in General Relativity, one can define a unique riemannian metric that interacts with everything (in this case the scalar field and its excitations) and characterizes a common background. 

The novelty of the present work is that the above mentioned result is actually general. In other words, for any non-linear scalar theory one can define a riemannian metric tensor which provides the geometrical structure ``seen" by the field. Therefore, there exist an effective spacetime ``generated'' by the non-linearity of the scalar field dynamics which will prescribe how this field propagates. A direct proof of our claim can be summarized in the following theorem.

\begin{theorem}\label{theorem}
Any scalar non-linear theory described by the lagrangian $L(w,\varphi)$ is equivalent to the field $\varphi$ propagating in an emergent spacetime with metric $\hat{h}_{\mu\nu}(\varphi,\partial\varphi)$ and a suitable source $j(\varphi, \partial\varphi)$, both constructed explicitly in terms of the field and its derivatives. Furthermore, in the optical limit, the wave vectors associated with its perturbations follow null geodesics in the same $\hat{h}_{\mu\nu}(\varphi,\partial\varphi)$ metric.
\end{theorem}
 
Proof: \\
The equation of motion (\ref{EqMotion}) describing the scalar field can be written as
\[
\frac{1}{\sqrt{-\gamma}}\partial_\mu \Bigl( \sqrt{-\gamma}\ L_{w} \, \partial_\nu \varphi \, \gamma^{\mu\nu} \Bigr) = \frac{1}{2}L_{\varphi} \qquad.
\]
We define the effective metric constructed with the lagrangian $L(w,\varphi)$, the Minkowskian metric $\gamma_{\mu \nu}$ and the scalar field $\varphi$
\begin{eqnarray}
\hat{h}_{\mu \nu} &\equiv& \frac{L_w}{\sqrt{1+\beta w}}\left(\gamma_{\mu \nu} -\frac{\beta}{1+\beta w} \varphi_{,\mu}\varphi_{,\nu}\right),\qquad \mbox{with} \quad \qquad \beta\equiv 2L_{ww}/L_{w}.
\end{eqnarray}
As a consequence of the  Cayley-Hamilton theorem, the determinant of a mixed tensor $\textbf{T}=T^{\alpha}_{\phantom a\beta}$ may be decomposed as a sum of traces of its powers in the form
\begin{eqnarray*}
det \textbf{T} &=& -\frac14 Tr\left(\textbf{T}^{4}\right) + \frac{1}{3} \,Tr \left(\textbf{T}\right) . Tr \left(\textbf{T}^{3}\right) + \frac{1}{8} Tr \left(\textbf{T}^{2}\right)^2 -\frac14 Tr\left(\textbf{T}\right)^2 . Tr \left(\textbf{T}^{2}\right)+\frac{1}{24} Tr \left(\textbf{T}\right)^4 \quad .
\end{eqnarray*}
Thus, the determinant of the $\hat h_{\mu\nu}$ is given by
\begin{equation}
\sqrt{-\hat{h}}=\frac{L_w^2}{\left(1+\beta w\right)^{3/2}}\sqrt{ -\gamma} \quad .
\end{equation}
Supposing that $L_{w}\neq 0$, the inverse is given through the relation $\hat{h}^{\mu \alpha}\hat{h}_{\alpha \nu}=\delta^\mu_{\, \nu}$, i.e.
\begin{eqnarray}\label{in}
\hat{h}^{\mu \nu} &\equiv& \frac{\sqrt{1+\beta w}}{L_w}\left(\gamma^{\mu \nu} +\beta \varphi^{,\mu}\varphi^{,\nu}\right) \quad .
\end{eqnarray}
Therefore, a straightforward calculation shows that
\begin{equation}
\hat{h}^{\mu\nu}\partial_{\nu}\varphi=\frac{\left(1+\beta w\right)^{3/2}}{L_w}\gamma^{\mu \nu}\partial_{\nu}\varphi \quad.
\end{equation}
Finally using the above relations, the equation of motion for the scalar field can be recast as
\begin{eqnarray}
\frac{1}{\sqrt{-\hat{h}}}\ \partial_{\mu}(\sqrt{-\hat{h}}\ \hat{h}^{\mu\nu}\partial_{\nu}\varphi)&=&\frac{L_{\varphi}}{2L_{w}^{2}}(1+\beta w)^{3/2} \quad .
\end{eqnarray}
Note that the left-hand side of this equation is nothing but the d'Alembertian constructed with the effective metric $\hat{h}_{\mu \nu}$. Also, the right hand-side does not involve more than the field and its first derivative. Using the same notation presented in relation (\ref{dal}) we obtain 
\begin{equation}\label{effEqMotion}
{\square}_{\hat{h}}\varphi = j(\varphi,\partial\varphi),
\end{equation}
where we have defined the effective source term as
\begin{equation}\label{effsource}
j(\varphi,\partial\varphi)\equiv \frac{L_{\varphi}}{2L_{w}^{2}}(1+\beta w)^{3/2} \qquad .
\end{equation}

The last step of our proof is straightforward once we realize that $\hat{h}_{\mu \nu}$ and $\hat{g}_{\mu \nu}$ are conformally related, indeed
\[
\hat{h}_{\mu \nu}=\frac{L_w^2}{\sqrt{1+\beta w}}\hat{g}_{\mu \nu} \quad .
\]
Recalling that equations (\ref{1}) and (\ref{2}) are conformally invariant, we can without further calculation write
\begin{eqnarray}
&&\hat{h}^{\alpha\beta}k_{\alpha}k_{\beta}=0 \\
&&\hat{h}^{\mu\nu}k_{\alpha ||\mu}k_{\nu}=0\quad ,
\end{eqnarray} 
and hence completing our proof showing that in the optical limit the wave vectors follow null geodesics in the $\hat{h}_{\mu\nu}$ metric.

Let us make some comments about what we have done. The novelty is that we have constructed an effective geometrical scenario to describe the dynamics of a nonlinear field and not just its perturbations. Thus, we are somehow generalizing previous results concerning analogue models of gravitation. In another way,  we can rephrased the above statement as follows: it is impossible to distinguish between a nonlinear field propagating in a Minkowski spacetime and the same field interacting minimally with an effective gravitational configuration $\hat{h}_{\mu\nu}$, constructed in terms of $\varphi$.

In the particular case of a theory where the lagrangian does not depend explicitly on $\varphi$, i.e. $L(w)$, equation (\ref{effEqMotion}) reduces to a ``free" wave propagating in a curved spacetime generated by itself 
\[\label{FREE}
{\square}_{\hat{h}}\varphi = 0 \quad.
\]

We should mention that the term free field has a peculiar meaning in this context. The effective metric is constructed with the scalar field $\varphi$, therefore, the above free Klein-Gordon equation is actually a complicated non-linear equation for $\varphi$. Notwithstanding, general relativity presents a very similar situation since besides the Klein-Gordon equation in curved spacetime, i.e. an intricate coupling between the metric and the scalar field, there is also Einstein's equations which describe how the scalar field modifies the spacetime metric. Thus, in GR the spacetime metric also depends on the scalar field in a non-trivial way. Of course, we can consider approximative situations where we truncate this back-reaction process and consider only the dynamics of the scalar field in a given spacetime that does not depends on its configuration. We shall examine this approximative situation in some details in section \ref{abri}.

Another point worth emphasizing is that it is the non-linearity in the kinetic term that produces the effective metric. If we consider algebraic non-linearities such that $ L(w,\varphi)=w+V(\varphi)$ with $V(\varphi)$ any function of the scalar field, the effective metric trivialize to the Minkowskian metric. Thus, it is the non-linearity in $w$ that it is essential to generate the curved effective spacetime.

Finally, to make connection with previous results concerning exceptional dynamics \cite{novello}, we mention that the unique lagrangian found in that work is recovered if one requires equation (\ref{in}) to be equal to (\ref{met}), which amounts to a differential equation for the lagrangian.

 
\section{Addressing the back-reaction issue}\label{abri}

Quantum field theory in curved spacetimes has been intensively investigated from the perspective of analogue gravity. The idea is to use a semi-classical approach where a physical situations can be approximated by a classical background field plus small quantum fluctuations satisfying linearized equations. Thus, the analogy only holds if we consider quantum effects where the gravitational back-reaction is negligible, i.e. the equations are essentially kinematics. An important discussion concerns the accurate description of these quantum fluctuations onto the dynamics of the classical background solution.  Although there exist some recent attempts to include these semi-classical back-reactions in the analogue gravity program (see, for instance {\cite{back}}), little has been said about this issue from its classical counterpart.

From the perspective of general relativity the back-reaction is basically the following. Classically, matter fields influence gravitation via its energy momentum tensor. Thus, any disturbance of the matter configuration immediately implies a modification of its background geometry. In this highly nonlinear process one has to take into account these altered metric back into the matter equations of motion. However, the situation is not so simple since the perturbed metric by itself depends on the perturbed field in a nontrivial way due to Einstein's equations. 

A very similar back-reaction process is included in our hidden metric perspective. To see how this works, let us suppose that we do know an exact continuous solution $\varphi_{0}$ of equation (\ref{effEqMotion}). The system behaves as a wave equation evolving in a metric $\hat{h}^{\mu\nu}_{0}$ (\ref{in}) with source $j_{0}$ (\ref{effsource}), both evaluated at $\varphi_{0}$. Now, suppose that we disturb this solution, i.e. we have a new scalar field and an associated metric in the form
\begin{eqnarray}\label{waveapprox}
&&\varphi_{1}=\varphi_{0}+\delta\varphi\\
&&\hat{h}^{\mu\nu}_{1}=\hat{h}^{\mu\nu}_{0}+\delta\hat{h}^{\mu\nu}
\end{eqnarray}
As a consequence of \textit{theorem 1}, the following equations results
\begin{equation}
{\square}_{\hat{h}_{0}}\varphi_{0} = j(\varphi_{0},\partial\varphi_{0})\quad, \qquad {\square}_{\hat{h}_{1}}\varphi_{1} = j(\varphi_{1},\partial\varphi_{1}).
\end{equation}
Thus, $\varphi_{0}$ and $\varphi_{1}$ propagates as waves associated to different metric structures. This happens because the disturbance of the background solution $\varphi_{0}$ implies a simultaneous disturbance of the background metric $\hat{h}^{\mu\nu}_{0}$.  

If we assume that the perturbations are infinitesimal, i.e. $\delta\varphi^{2}<<\delta\varphi$, we obtain, up to first order, the linear equation 
\begin{equation}\label{linea}
\left[\delta\left(\sqrt{-\hat h}\ \hat h^{\mu\nu}\right)\varphi_{0,\nu}+\sqrt{-\hat h_{0}}\ \hat h^{\mu\nu}_{0}\delta\varphi_{,\nu}  \right]_{,\mu}=j_{0}\ \delta\sqrt{-\hat h}+\sqrt{-\hat h_{0}}\ \delta j
\end{equation}
where all the background quantities depend only on position and all the perturbed quantities $\delta j$ and $\delta \hat{h}^{\mu\nu}$ are to be written in terms of $\delta\varphi$ and its derivatives $\delta\varphi_{,\alpha}$, i.e.
\begin{equation} 
\delta\hat h^{\mu\nu}=\frac{\partial \hat h^{\mu\nu}}{\partial \varphi}\delta\varphi+\frac{\partial \hat h^{\mu\nu}}{\partial \varphi_{,\alpha}}\delta\varphi_{,\alpha},\quad\quad
\delta j=\frac{\partial j}{\partial \varphi}\delta\varphi+\frac{\partial j}{\partial \varphi_{,\alpha}}\delta\varphi_{,\alpha}.
\end{equation}
Using the usual relation between the variation of the determinant and the variation of the metric
\begin{equation}
\delta\sqrt{-\hat h}=-\frac{1}{2}\sqrt{-\hat h}\ \hat h_{\alpha\beta}\ \delta\hat h^{\alpha\beta} 
\end{equation}
we obtain, after a tedious but straightforward calculation, that equation (\ref{linea}) is simply the well known  Klein-Gordon-like equation in a given curved background (\ref{wavv})
\begin{equation}
\square_{\hat{f}}\delta\varphi+m^{2}\delta\varphi=0
\end{equation}
where both the metric $\hat{f}^{\mu \nu}$ and the mass term $m^2$ are calculated with respect to the $\varphi_0$ configuration. This expression is exactly the same as that discussed in section \ref{waveprop}, which is by no means a coincidence. Once eq. (\ref{effEqMotion}) is completely equivalent to (\ref{EqMotion}), the perturbations (\ref{waveapprox}) must coincide with the previous perturbative approach (\ref{pertu}). Nevertheless, note that in order to obtain this last result, it is indispensable to include the variation of the background metric. This characterizes a process that is similar to a gravitational back-reaction. The background metric $\hat{f}_{\mu\nu}$ that rules the motion of perturbations $\delta\varphi$ may be obtained in terms of the metric $\hat{h}_{\mu\nu}$ that rules the motion of the whole field $\varphi$ evaluated at the background configuration, up to first order as
\begin{equation}\label{gama}
{\hat{f}}^{\mu\nu}=\left[(1+\beta w)^{-1}\hat{h}^{\mu\nu}\right]_{\varphi_0}
\end{equation}

It is interesting to notice that $\hat{f}^{\mu \nu}$ and $\hat{h}^{\mu \nu}$ are related by a simple conformal transformation. We thus recover all the previous linearized results concerning field theory in curved spacetimes. In the limit of a wave-length sufficiently small we re-obtain also the eikonal approximation discussed in section \ref{rayprop}.

\section{Example: The geometry of hydrodynamic flows}\label{ghf}

It is well known that any theory of the form $L(w,\varphi)$ with a timelike gradient $\partial_{\mu}\varphi$ ($w>0$) may be alternatively described as an effective hydrodynamic flow \cite{vikman}-\cite{outra}. In this section we would like to apply our method to describe such a relativistic fluid. We restrict ourselves to the case of an irrotational barotropic flow which has only one degree of freedom i.e. can be described by a single scalar field. The particular situation where the lagrangian does not depend explicitly on $\varphi$ gives rise to a hydrodynamical flow with conserved number of particles \cite{taub},\cite{schutz2} and \cite{outra}. We will investigate this simplified configuration in what follows.

First,  note that the energy momentum tensor of a scalar field with a lagrangian that is not an explicit function of $\varphi$, i.e. $L(w)$ is given by 
\begin{equation}\label{energy}
T_{\mu\nu}\equiv \frac{2}{\sqrt{-\gamma}}\frac{\delta\sqrt{-\gamma}L}{\delta \gamma^{\mu\nu}}=       2L_{w}\varphi_{,\mu}\varphi_{,\nu}-L\gamma_{\mu\nu} \quad.
\end{equation}

Assuming that $w>0$ we can define a normalized timelike congruence of observers comoving with the fluid 
\begin{equation}
v_{\mu}=\frac{\partial_{\mu}\varphi}{\sqrt{w}}
\end{equation}
the vorticity of which $w_{\alpha\beta} \equiv v_{[\mu \, ; \nu]}$ is identically zero. Thus, we identify the scalar $\varphi$ with a velocity potential. Furthermore, the timelike constraint on $\partial_{\mu}\varphi$ implies that the anisotropic pressure $\pi_{\mu\nu}$ and the heat-flux vector $q_{\mu}$ identically vanish. Thus, the energy momentum tensor (\ref{energy}) describes a perfect fluid with energy density $\rho$ and pressure $p$ given by the relations
\begin{equation}\label{press}
\rho=2L_{w}w-L,\quad\quad p=L.
\end{equation}

Note that it is possible to write the pressure as a function of the energy density, thus yielding a barotropic equation of state given by $p(\rho)$. We define for future convenience the velocity of sound, particle density and enthalpy respectively as 
\begin{eqnarray}\label{thermorel}
c_{s}^{2}\equiv\frac{\partial p}{\partial\rho}=\frac{1}{1+\beta w}\quad , \qquad n\equiv exp\int\frac{d\rho}{\rho+p}=\sqrt{w}L_{w}\quad ,\qquad \mu\equiv\frac{\rho+p}{n}=2\sqrt{w} \quad .
\end{eqnarray}

The dynamical equations governing the motion of a perfect fluid are immediately obtained by projecting the conservation law $T^{\mu\nu}_{\phantom a\phantom a ;\nu}=0$
\begin{eqnarray*}
&&\dot{\rho}+(\rho+p)\theta=0\\\\ 
&&p_{,\alpha}\, P^{\alpha}_{\phantom a\mu}+(\rho+p)a_{\mu}=0
\end{eqnarray*}
where $P_{\mu\nu}\equiv g_{\mu\nu}-v_{\mu}v_{\nu}$ is the projector, $\theta\equiv v^{\mu}_{\phantom a ;\mu}$ the expansion scalar and $a_{\mu}\equiv v_{\mu ;\nu}v^{\nu}$ the acceleration four-vector. In the case under investigation all the physical quantities that appear in the above equations depend explicitly on the field $\varphi$ and its derivatives. Thus, these equations are not independent. On the other side, calculating explicitly the divergence of the energy-momentum tensor directly from its definition (\ref{energy}) we obtain
 \begin{equation}
T^{\mu\nu}_{\phantom a\phantom a ;\nu}=\varphi^{,\mu}(L_{w}\gamma^{\alpha\beta}\varphi_{,\alpha})_{;\beta}=0 \quad.
\end{equation}

Thus, the conservation of the fluid's energy-momentum is nothing but equation (\ref{EqMotion}) with $L_{\varphi}=0$. Consequently, our theorem guarantees that the fluid flow may be described as an effective wave equation of the form $\square_{\hat{h}}\varphi=0$. Using relations (\ref{thermorel}) the effective metric finally reads
\begin{equation}
\hat{h}^{\mu\nu}=\frac{\mu}{2nc_{s}}\left[\gamma^{\mu\nu}+(c_{s}^{-2}-1)v^{\mu}v^{\nu}\ \right] \quad.
\end{equation}

Summarizing:

\begin{itemize}
\item{the dynamics of a relativistic perfect fluid with a barotropic equation of state is such that its velocity potential evolves as a wave embedded in an emergent curved manifold generated by the wave itself.}
\end{itemize}

A similar metric was first obtained by Moncrief \cite{moncrief} in studying spherical accretion of matter onto a non-rotating black hole. Recently, a similar result was obtained by Vikman in the context of k-essence theories \cite{vikman}. Nevertheless, both investigations were restricted to the case of non-gravitating fluid perturbations, i.e. acoustic propagation. In our scheme, all these previous results may be obtained using the metric $\hat{f}^{\mu\nu}$ instead of $\hat{h}^{\mu\nu}$. Here, the effective spacetime governs not only the sound cone of fluid excitations (characteristic surfaces) but also fluid dynamics itself.

Note that the converse is also true. Any fluid may be mapped into a lagrangian of the form $L(w)$. To see what lagrangian corresponds to a given $p=f(\rho)$ it is convenient to invert the barotropic equation to obtain $\rho=f^{-1}(p)$. Now, relations (\ref{press}) implies the differential equation
\begin{equation}
 2w\frac{dL}{dw}=L+f^{-1}(L) \quad .
\end{equation}
For a given fluid, in general, there is a lagrangian $L(w)$ which is a solution of the above equation. In this scenario, fluids with constant equation of state which are very popular in cosmology, i.e. $p=\lambda \rho$ with $\lambda$ constant,  are simply power law solutions 
\begin{eqnarray*}
L(w)&=&\frac{2\lambda}{1+\lambda}w^{\left(1+\lambda\right)/2\lambda}\qquad \Rightarrow \qquad p=\lambda \rho,
\end{eqnarray*}
which gives rise to the effective metric
\begin{equation}
\hat{h}^{\mu\nu}_{(\lambda)}=\frac{\kappa_{\lambda}}{\rho^{(1-\lambda)/\left(1+\lambda\right)}}\left[\gamma^{\mu\nu}+\frac{\alpha_\lambda}{\rho^{2\lambda/(1+\lambda)}}v^\mu v^\nu\right]
\end{equation}
with
\begin{eqnarray*}
\kappa_\lambda=\frac{1}{\sqrt{\lambda}}\left(\frac{1+\lambda}{2}\right)^{(1-\lambda)/(1+\lambda)}\quad &,&\quad \alpha_\lambda=\frac{1-\lambda}{\lambda}\left(\frac{2}{1+\lambda}\right)^{2\lambda/(1+\lambda)}\quad.
\end{eqnarray*}

This last result implies that the scalar field $\varphi$ governing the flow of any perfect fluid may be entirely described in terms of an effective metric generated by itself. It would be interesting to apply our scheme to the study of these fluids in the context of cosmology. We shall exemplify with a simplified newtonian limit described in the appendix.



 
\section{Conclusion}

In this paper we have investigated the relation between the equation of motion of a relativistic self-interacting field with a wave propagating in a curved spacetime. We have shown that for any theory of the form $L(w,\varphi)$ there always exist a spacetime endowed with a metric $\hat h^{\mu\nu}$ such that both above mentioned dynamics are equivalent. Hence, the dynamics of the non-linear theory can be described as a minimal coupling with an emergent gravitational metric constructed with the scalar field and its derivatives. The novelty of our analysis is that geometrization is an universal process, i.e. it is valid for an arbitrary lagrangian. As a concrete application, due to the fact that there is a formal equivalence between a scalar field and an ideal barotropic fluid, we used our geometrization scheme to describe the evolution of an irrotational hydrodynamical flow.


As long as our result does not rely on any kind of approximation, typical to effective metric techniques, we hope that it can shed light and open new perspectives for the analogue gravity program. In particular, it would be very interesting to investigate further consequences of our formalism in the intricate semi-classical back-reaction or field quantization.

We suspect that our analysis might be generalized to describe other kinds of nonlinear fields such as spinors, vector or tensors. We will come back to this discussion elsewhere.

\section{Appendix I: Hydrodynamics with Newtonian Approximation }

 Analogue models with fluids in the non-relativistic regime has been extensively studied due to its valuable applications. As a typical example, there is the case of the so-called acoustic black holes \cite{Un}, \cite{Vi} in which the sound wave disturbances in an accelerated fluid mimic the propagation of light in a curved space-time. When the velocity of the background flow reaches the speed of sound characteristic of this medium a sonic horizon is created trapping the sound just as gravitational black hole traps light inside the event horizon. These models have opened a promising avenue to study important aspects of physics in curved spacetimes such as Hawking's radiation or some phenomenological corrections coming from quantum gravitation \cite{Na}.

In order to find a newtonian approximation of our results presented in Section V, we should take the limit where the velocity field $v^\mu$ goes to $v^\mu=(1, \vec v)$, where $|\vec v| << 1$, and the energy density is much higher than the pressure so that $n\rightarrow\rho$ and $\mu\rightarrow 1$. In this case, the effective metric reads,
\begin{equation}
	\hat{h}^{\mu\nu}\rightarrow\frac{1}{2\rho c_s}\left(c_s^2\eta^{\mu\nu} + v^\mu v^\nu\right),
\end{equation}
but, as shown before, excitations of the fluid (phonons) propagate along null geodesics in a conformal metric given by equation (\ref{gama}), i.e.
\begin{eqnarray}
	\hat {f}^{\mu\nu}=\frac{1}{2\rho c_s}\left(\begin{array}{ccc}
1 & \vdots & v^j\\
\cdots\cdots\cdots & \cdot & \cdots\cdots\cdots\cdots\cdots \\
v^i & \vdots & (-c^{2}_{s}\delta^{ij}+ v^i v^j)\end{array}\right),
\end{eqnarray}
this is the same result obtained in that models (up to some definitions such as the space-time signature). Therefore, from our geometrization of hydrodynamics, we can also get the usual non-relativistic fluid models.

\section{acknowledgements}
M. Novello would like to thank FINEP, CNPq and FAPERJ, E. Goulart FAPERJ and J. D. Toniato CNPq for their financial support.


\end{document}